# On the Value of Correlation

Itai Ashlagi    Dov Monderer*    Moshe Tennenholtz†

Faculty of Industrial Engineering and Management
Technion–Israel Institute of Technology
Haifa 32000, Israel

## Abstract

Correlated equilibrium (Aumann, 1974) generalizes Nash equilibrium to allow correlation devices. Aumann showed an example of a game, and of a correlated equilibrium in this game, in which the agents' surplus (expected sum of payoffs) is greater than their surplus in all mixed-strategy equilibria. Following the idea initiated by the price of anarchy literature (Koutsoupias & Papadimitriou, 1999; Papadimitriou, 2001) this suggests the study of two major measures for the value of correlation in a game with non-negative payoffs:

1. The ratio between the maximal surplus obtained in a correlated equilibrium to the maximal surplus obtained in a mixed-strategy equilibrium. We refer to this ratio as the mediation value.
2. The ratio between the maximal surplus to the maximal surplus obtained in a correlated equilibrium. We refer to this ratio as the enforcement value.

In this work we initiate the study of the mediation and enforcement values, providing several general results on the value of correlation as captured by these concepts. We also present a set of results for the more specialized case of congestion games (Rosenthal, 1973), a class of games that received a lot of attention in the recent literature.

## 1 Introduction

One of the most famous and fruitful contributions to game theory has been the introduction of correlated equilibrium by Aumann (1974) . Consider a game in strategic form. A correlated strategy is a probability distribution over the set of strategy profiles, where a strategy profile is a vector of strategies, one for each player. A correlated strategy is utilized as follows: A strategy profile is selected according to the distribution, and every player is informed about her strategy in the profile. This selected strategy for the player is interpreted as a recommendation of play. Correlated strategies are most natural, since they capture the idea of a system administrator/reliable party who can recommend behavior but can not enforce it. Hence, correlated strategies make perfect sense in the context of congestion control, load balancing, trading, etc. A correlated strategy is called a correlated equilibrium if it is better off for every player to obey her recommended strategy if she believes that all other players obey their recommended strategies [1]. A major potential benefit of correlated equilibrium is to attempt to improve the social welfare of selfish players. In this paper, the social welfare obtained in a mixed-strategy profile is defined to be the expected sum of the payoffs of the players, and it is referred to as the surplus obtained in this profile.

A striking example introduced in Aumann's seminal paper (Aumann, 1974) is of a two-player two-strategy game, where the surplus obtained in a correlated equilibrium is higher than the surplus obtained in every mixed-strategy equilibrium of the game. As a result, Aumann's example suggests that correlation may be a way to improve upon social welfare while still assuming that players are rational in the classical game-theoretic sense.[2]

*Supported in part by grant number 1380001 from the Israel Science Foundations (Bikura).

†Supported in part by grant number 53/03-10.5 from the Israel Science Foundations.

---

[1] Every correlated strategy defines a Bayesian game, in which the private signal of every player is her recommended strategy. It is a correlated equilibrium if obeying the recommended strategy by every player is a pure-strategy equilibrium in the Bayesian game.

[2] Other advantages are purely computational ones. As has been recently shown correlated equilibrium can be computed in polynomial time even for structured representa-

A modification of Aumann's example serves us as a motivating example:

|       | $b^1$ | $b^2$ |
|-------|-------|-------|
| $a^1$ | 5, 1  | 0, 0  |
| $a^2$ | 4, 4  | 1, 5  |

**Aumann's Example**

In this game, there are three mixed-strategy equilibrium profiles. Two of them are obtained with pure strategies, $(a^1, b^1)$, and $(a^2, b^2)$. The surplus in each of these pure-strategy equilibrium profiles equals six. There is an additional mixed-strategy equilibrium in which, every player chooses each of her strategies with equal probabilities. The surplus obtained in this profile equals 5 ($= \frac{1}{4}(6+0+8+6)$) because every entry in the matrix is played with probability $\frac{1}{4}$. Hence, the maximal surplus in a mixed-strategy equilibrium equals 6. Consider the following correlated strategy: a probability of 1/3 is assigned to every pure strategy profile but $(a^1, b^2)$. This correlated strategy is a correlated equilibrium. Indeed, when the row player is recommended to play $a^1$ she knows that the other player is recommended to play $b^1$, and therefore she strictly prefers to play $a^1$. When the row player is recommended to play $a^2$ the conditional probability of each of the columns is half, and therefore she weakly prefers to play $a^2$. Similar argument applied to the column player shows that the correlated strategy is indeed a correlated equilibrium. The surplus associated with this correlated equilibrium equals $\frac{20}{3}$ ($= \frac{1}{3}(6+8+6)$).

The above discussion suggests one may wish to consider the value of correlation in games. In order to address the challenge of studying the value of correlation, we tackle two fundamental issues:

- How much can the society/system gain by adding a correlation device, where we assume that without such a device the agents play a mixed strategy equilibrium.

- How much does the society/system loose from the fact that the correlation device can only recommend (and can not enforce) a course of action?

We introduce two measures, namely the *mediation value* and the *enforcement value*. The mediation value will measure the ratio between the maximal surplus in a correlated equilibrium to the maximal surplus in a mixed-strategy equilibrium. Notice that the higher

___
tions of games Kakade *et al.* (2003); Papadimitriou (2005).

this number is, the more correlation helps. This concept relates to the price of anarchy[3] as follows: When translating the definition of price of anarchy to games with payoffs and not with costs,[4] the price of anarchy is the ratio between the maximal surplus to the minimal surplus obtained in a mixed-strategy equilibrium. The higher this number is, the value of a center is higher, where a center can enforce a course of play. Hence, the price of anarchy could have been called the value of a center with respect to anarchy, where a center can dictate a play, and when anarchy is measured by the worst social outcome reached by rational and selfish agents. The mediation value is the value of a center with respect to anarchy, where a center is reliable and can recommend a play, and anarchy is measured by the best social outcome reached by rational and selfish agents.[5]

In Aumman's example it can be shown that the correlated equilibrium introduced above is the best correlated equilibrium, i.e. it attains the maximal surplus among all correlated equilibria in the game. Hence, the mediation value of Aumann's game is $\frac{10}{9}$.

The enforcement value measures the ratio between the maximal surplus to the maximal surplus in a correlated equilibrium. That is, it is the value of a super-center with respect to a center who can just use correlated devices in equilibrium. As the maximal surplus in Aumann game is 8, the enforcement value in this game equals $\frac{6}{5}$.

In order for the above measures to make sense we consider games with non-negative payoffs.

In this paper we establish general and basic results concerning the measures defined above. We will consider the mediation ( enforcement) value of classes of games, where the mediation ( enforcement) value of a class of games is defined as the greatest lower bound of the mediation (enforcement) values of the games in the class.

We start by considering general games. Aumann's ex-ample implies that the mediation value of the class of two-player two-strategy ($2 \times 2$) games is at least $10/9$. We first show that the mediation value of this class is $4/3$. Hence, the mediation value in any $2 \times 2$ game is bounded from above by $4/3$, and this upper bound is tight. Next we move to show the power of correlation in more complex games. In order to do so we consider the two possible minimal extensions of $2 \times 2$ games: Two-player games with three strategies for one of the players and two strategies for the other, and three-player games with two strategies for all players. We show that the mediation values of the games in each of these classes of games are unbounded. That is, the mediation value equals $\infty$. This implies that the mediation value is $\infty$ for the class of games in which, at least one agent has three strategies and for the class of games with at least three players. Again, this should be interpreted as a positive result, showing the extreme power of correlation.

___
[3]The concept of the price of anarchy has received much attention in the recent computer science literature. See e.g., (Czumaj & Vocking, 2002; Marvonicolas & Spirakis, 2001; Roughgarden, 2002; Roughgarden & Tardos, 2002).

[4]In many situations it is indeed natural to deal with non-negative costs rather than payoffs. Such models require special treatment. Interestingly, it can be shown that there are classes of games where the price of anarchy is bounded when dealing with costs while it is unbounded when dealing with payoffs. Similarly for the mediation value. This point will be further discussed in the full paper.

[5]Anarchy means playing in a mixed-strategy equilibrium. The phenomenon of multiple equilibria forces a modelling choice. Currently the choice between the best and worst social outcomes is a matter of taste.

Considering the enforcement value, we first show that it equals $\infty$ for the class of $2 \times 2$ games. Moreover, in a setup with three players we show that the enforcement value of the class of three-player games without dominated strategies equals $\infty$.

Following these general results, we consider the important class of congestion games (Rosenthal, 1973; Monderer & Shapley, 1996). Indeed, this class of games is perhaps the most applicable to the game theory and CS synergy. In particular, results regarding the price of anarchy have been obtained for congestion games. We restrict our discussion to simple congestion games. In a simple congestion game there is a set of facilities. Every facility $j$ is associated with a payoff function $w_j$. Every player chooses a facility, say facility $j$, and receives $w_j(k)$, where $k$ is the number of players that chose facility $j$.

For completeness we first deal with the simple case where we have only two players. In this case we show that if the players can choose among only two facilities then the mediation value is bounded from above by $4/3$, and that this bound is achieved. In the more general case, where there are $m$ facilities, the mediation value is bounded by $2$. However, if we consider facilities with non-increasing payoffs (i.e. a player's payoff is monotonically non-increasing in the number of agents using its selected facility) then the mediation value is $1$.

We then move to the more general case of simple congestion games, where there are $n \geq 2$ players. We show that for the case of three players, even if there are two facilities with non-increasing payoffs then the mediation value is unbounded. However, if we have $n$ players and the two facilities have non-increasing *linear* payoff functions then the mediation value is bounded from above by $\frac{\sqrt{5}+1}{2}$. On the other hand, we show that the mediation value can be higher than $1$ considering even two symmetric (identical) facilities with non-increasing payoffs. This further illustrates the power of correlation. Nevertheless, we also show that if we have $m$ symmetric facilities, where the facility payoff functions obey a concavity requirement, the best mixed-strategy equilibrium obtains the maximal surplus, and therefore both the mediation value and the enforcement value are $1$.

Finally, we study the enforcement value in the natural case where we have $n$ players who choose among $m$ symmetric facilities (where the payoff function associated with a facility may be arbitrary). We give a general characterization of the cases where the enforcement value is $1$, and as a result determine the situations where correlation allows obtaining maximal surplus. Few other results about the enforcement value are obtained as well.

Most of our proofs are omitted from this paper due to lack of space. The proofs rely mainly on duality theorems in linear programming. In order to illustrate these techniques we added a short appendix with a discussion of these techniques and a short sketch of proof of one of our theorems.

## 2 Basic Definitions

A finite game $\Gamma = (N, (S^i)_{i \in N}, (u^i)_{i \in N})$ in strategic form is defined as follows. Let $N$ be a nonempty finite set of players. For each $i \in N$, let $S^i$ be a finite set of strategies of player $i$. Let $S = S^1 \times S^2 \times \cdots \times S^n$ be the set of strategy profiles (n-tuples). An element of $S$ is $s = (s^i)_{i \in N}$. For each $i \in N$ and $s \in S$ let $s^{-i} = (s^1, ..., s^{i-1}, s^{i+1}, ...s^n)$ denote the strategies played by everyone but $i$. Thus $s = (s^{-i}, s^i)$. For each player $i \in N$, let $u^i : S \to \mathbf{R}$ be the payoff function of player $i$. $u^i(s)$ is the payoff of player $i$ when the profile of strategies $s$ is played. $\Gamma$ is called a *nonnegative game* if all payoffs to all players are nonnegative, i.e $u^i : S \to \mathbf{R}_+$.

A player can also randomize among her strategies by using a mixed strategy - a distribution over her set of strategies. For any finite set $C$, $\Delta(C)$ denotes the set of probability distributions over $C$. Thus $P^i = \Delta(S^i)$ is the set of mixed strategies of player $i$. For every $p^i \in P^i$ and every $s^i \in S^i$, $p^i(s^i)$ is the probability that player $i$ plays strategy $s^i$. Every strategy $s^i \in S^i$ is, with the natural identification, a mixed strategy $p_{s^i} \in P^i$ in which

$$p_{s^i}(t^i) = \begin{cases} 1 & t^i = s^i \\ 0 & t^i \neq s^i \end{cases}.$$

$p_{s^i}$ is called a *pure* strategy, and $s^i$ is interchangeably called a strategy and a pure strategy (when it is identified with $p_{s^i}$). Let $P = P^1 \times P^2 \times \cdots \times P^n$ be the set of mixed strategy profiles.

Unless otherwise specified we will assume that $N = \{1, 2, ...., n\}$, $n \geq 1$.

Any $\mu \in \Delta(S)$ is called a *correlated strategy*. Every mixed strategy profile $p \in P$ can be interpreted as a correlated strategy $\mu_p$ in the following way. For every strategy profile $s \in S$ let $\mu_p(s) = \prod_{i=1}^n p^i(s^i)$. With slightly abuse of notation, for every $\mu \in \Delta(S)$, we denote by $u^i(\mu)$ the expected payoff of player $i$ when the correlated strategy $\mu \in \Delta(S)$ is played, that is:

$$u^i(\mu) = \sum_{s \in S} u^i(s)\mu(s). \qquad (1)$$

Whenever necessary we identify $p$ with $\mu_p$. Naturally, for every $p \in P$ let $u^i(p) = u^i(\mu_p)$. Hence $u^i(p)$ is the expected payoff of player $i$ when the mixed strategy $p$ is played.

We say that $p \in P$ is a *mixed-strategy equilibrium* if $u^i(p^{-i}, p^i) \geq u^i(p^{-i}, q^i)$ for every player $i \in N$ and for every $q^i \in P^i$.

**Definition 2.1** *(Aumann 1974, 1987) A correlated strategy $\mu \in \Delta(S)$ is a correlated equilibrium of $\Gamma$ if and only if for all $i \in N$ and all $s^i, t^i \in S^i$:*

$$\sum_{s^{-i} \in S^{-i}} \mu(s^{-i}, s^i)[u^i(s^{-i}, s^i) - u^i(s^{-i}, t^i)] \geq 0. \quad (2)$$

It is well-known and easily verified that every mixed-strategy equilibrium is a correlated equilibrium. Let $u(\mu) = \sum_{i=1}^n u^i(\mu)$. The value $u(\mu)$ is called the *surplus* at $\mu$. Let $N(\Gamma)$ be the set of all mixed-strategy equilibria in $\Gamma$ and let $C(\Gamma)$ be the set of all correlated equilibria in $\Gamma$. We define $v_C(\Gamma)$ and $v_N(\Gamma)$ as follows:

$$v_C(\Gamma) \triangleq \max\{u(\mu) : \mu \in C(\Gamma)\},$$

$$v_N(\Gamma) \triangleq \max\{u(p) : p \in N(\Gamma)\}.$$

Note that $v_N(\Gamma)$ and $v_C(\Gamma)$ are well defined due to the compactness of $N(\Gamma)$ and $C(\Gamma)$ respectively, and the continuity of $u$. Define $opt(\Gamma)$ (the maximal surplus) as follows:

$$opt(\Gamma) \triangleq \max\{u(\mu) : \mu \in \Delta(S)\} = max\{u(s) : s \in S\}.$$

The *mediation value* of a nonnegative game $\Gamma$ is defined as follows:

$$MV(\Gamma) \triangleq \frac{v_C(\Gamma)}{v_N(\Gamma)}.$$

If both $v_N(\Gamma) = 0$ and $v_C(\Gamma) = 0$ we define $MV(\Gamma)$ to be 1. If $v_N(\Gamma) = 0$ and $v_C(\Gamma) > 0$ then $MV(\Gamma)$ is defined to be $\infty$. Denote by $EV(\Gamma)$ the *enforcement value* of a nonnegative game $\Gamma$. That is,

$$EV(\Gamma) \triangleq \frac{opt(\Gamma)}{v_C(\Gamma)}.$$

If both $v_C(\Gamma) = 0$ and $opt(\Gamma) = 0$ then we define $EV(\Gamma)$ to be 1. If $v_C(\Gamma) = 0$ and $opt(\Gamma) > 0$ then $EV(\Gamma)$ is defined to be $\infty$. Finally, for a class of games $\mathcal{C}$ we denote

$$MV(\mathcal{C}) \triangleq \sup_{\Gamma \in \mathcal{C}} MV(\Gamma); \quad \text{and} \quad EV(\mathcal{C}) \triangleq \sup_{\Gamma \in \mathcal{C}} EV(\Gamma).$$

We will also make use of the following notation and definitions. Let $\mathcal{G}$ be the class of all nonnegative games in strategic form. For $m_1, m_2, ..., m_n \geq 1$ denote by $\mathcal{G}_{m_1 \times m_2 \times \cdots \times m_n} \subseteq \mathcal{G}$ the class of all games with $n$ players in which $|S^i| = m_i$ for every player $i$. Let $s^i, t^i \in S^i$ be pure strategies of player $i$. We say that $s^i$ *weakly dominates* (or just dominates) $t^i$, and $t^i$ is *weakly dominated* (or dominated) by $s^i$ if for all $s^{-i} \in S^{-i}$

$$u^i(s^i, s^{-i}) \geq u^i(t^i, s^{-i}),$$

where at least one inequality is strict. We say that $s^i$ *strictly dominates* $t^i$, and $t^i$ is *strictly dominated* by $s^i$ if all of the above inequalities are strict. If $u^i(s^i, s^{-i}) = u^i(t^i, s^{-i})$ for all $s^{-i} \in S^{-i}$ then we will say that $s^i$ and $t^i$ are *equivalent* strategies for player $i$.

## 3 Results for General Games

We now deal with general games in strategic form.

### 3.1 The Mediation Value

In this section we show the overwhelming power of correlation in general games. However, we start with extending Aumann's result on the power of correlation in $2 \times 2$ games.

#### 3.1.1 Two-person two-strategy games

Aumann's example shows that a mediation value of $\frac{10}{9}$ can be obtained in a $2 \times 2$ game. In order to study the value of correlation we prove:

**Theorem 3.1** $MV(\mathcal{G}_{2 \times 2}) = \frac{4}{3}$.

We now show a family of games in which the mediation value approaches the above $\frac{4}{3}$ bound. Consider the family of games $\Gamma_x$ shown in Figure 1 (a variant of Aumann's example) where $x > 1$.

|       | $b^1$   | $b^2$ |
|-------|---------|-------|
| $a^1$ | $x$ \ 1 | 0 \ 0 |
| $a^2$ | $x-1$ \ $x-1$ | 1 \ $x$ |

Figure 1

In this game the pure strategy profiles $(a^1, b^1)$ and $(a^2, b^2)$ are in equilibrium and $u(a^1, b^1) = u(a^2, b^2) = x + 1$. There is one more equilibrium in mixed strategies where each player assigns the probability 0.5 to each of her strategies, which yields a surplus lower than $x + 1$. The correlated strategy $\mu \in \Delta(S)$ where each of the strategy profiles $(a^1, b^1), (a^2, b^1)$ and $(a^2, b^2)$ is played with equal probability $1/3$ is in equilibrium and $u(\mu) = \frac{4x}{3}$. $\mu$ obtains the largest surplus among all correlated equilibria in the game (see the proof). Hence $MV(\Gamma_x) = \frac{4x}{3(x+1)}$. Therefore $MV(\Gamma_x) \to \frac{4}{3}$ when $x \to \infty$. $\square$

#### 3.1.2 General Games

The above theorem shows that the mediation value of the class of $2 \times 2$ games is finite. A major question we face is whether such finite bound exists for more general classes of games. We now show that, perhaps surprisingly, the mediation value equals $\infty$ if we consider slightly more complex games. In particular, if we allow one of the players in a 2-player game to have at least three strategies, while the other remains with two strategies, then the mediation value already equals $\infty$. Similarly, if we allow three players each with two strategies then the mediation value, again equals $\infty$. Together, these results show the power of correlation when we move beyond $2 \times 2$ games.[6]

**Theorem 3.2** $MV(\mathcal{G}_{m_1 \times m_2}) = \infty$ for every $m_1, m_2 \geq 2$ such that $\max(m_1, m_2) \geq 3$.

**Theorem 3.3** $MV(\mathcal{G}_{m_1 \times \cdots \times m_n}) = \infty$ for every $n \geq 3$ and for every $m_1, m_2, \cdots, m_n \geq 2$.

### 3.2 The Enforcement Value

We first show that the enforcement value may be unbounded even on classes of small games.

**Theorem 3.4** $EV(\mathcal{G}_{m_1 \times \cdots \times m_n}) = \infty$ for every $n \geq 2$, and for every $m_1, m_2 \geq 2$.

---

[6]The results presented in this paper showing that the mediation value may be $\infty$, can also be established when we assume that the payoffs are uniformly bounded, e.g. when all payoffs are in the interval $[0, 1]$. This fact, which strengthen our results even further, will be discussed in the full paper.

The proof of Theorem 3.4 is shown using a parametric version of the well-known Prisoner's Dilemma game. This game has the property of possessing a strictly dominant strategy for each player. In the next theorem we show that dominance is not necessary for obtaining an unbounded enforcement value.

**Theorem 3.5** $\sup\{EV(\Gamma)|\Gamma \in \mathcal{G}_{2 \times 2 \times 2},$
*no player has a strictly dominant strategy*$\} = \infty$.

## 4 Simple Congestion Games

In this section we explore the mediation and enforcement values in simple congestion games. We first need a few notations and definitions.

A congestion form $F = (N, M, (X^i)_{i \in N}, (w_j)_{j \in M})$ is defined as follows. $N$ is a nonempty set of players and $M$ is a nonempty set of facilities. Unless otherwise specified we let $M = \{1, 2, ..., m\}$. For $i \in N$, let $X^i$ be the set of strategies of player $i$, where each $A^i \in X^i$ is a nonempty subset of $M$. For $j \in M$ let $w_j \in R^{\{1,2,...,n\}}$ be the facility payoff function, where $w_j(k)$ denotes the payoff of each user of facility $j$, if there are exactly $k$ users. A congestion form is nonnegative if for every $j \in M$ $w_j$ is nonnegative. A congestion form is *simple* if for every $i \in N$, $X^i = \{\{1\}, \{2\}, ..., \{m\}\}$. Let $\mathcal{S}$ be the class of all nonnegative simple congestion forms and denote by $\mathcal{S}_{n \times m} \subseteq \mathcal{S}$ the class of all nonnegative simple congestion forms with $n$ players and $m$ facilities. Every congestion form $F = (N, M, (X^i)_{i \in N}, (w_j)_{j \in M})$ defines a *congestion game* $\Gamma_F = (N, (X^i)_{i \in N}, (u^i)_{i \in N})$ where $N$ and $X^i$ are as above and $(u^i)_{i \in N}$ is defined as follows.

Let $X = \times_{i \in N} X^i$. For every $A = (A^1, A^2, ..., A^n) \in X$ and every $j \in M$ let $\sigma_j(A) = |\{i \in N : j \in A^i\}|$ be the number of users of facility $j$.
Define $u^i : X \to \mathbb{R}$ by

$$u^i(A) = \sum_{j \in A^i} w_j(\sigma_j(A)). \quad (3)$$

Observe that if $F$ is simple then $u^i(A) = w_{A^i}(\sigma_{A^i}(A))$.

We will say that a facility $j$ is *non-increasing* if $w_j(k)$ is a non-increasing function of $k$. Define $\mathcal{SN}_{n \times m} \subseteq \mathcal{S}_{n \times m}$ as follows:

$\mathcal{SN}_{n \times m} \triangleq$
$\{F \in \mathcal{S}_{n \times m} | \text{all facilities in F are non-increasing}\}$.

We will call a facility $j$ *linear* if there exist a constant $d_j$ such that $w_j(k+1) - w_j(k) = d_j$ for every $k \leq 1$.

A congestion form is called *facility symmetric* or just *symmetric* if $w_j \equiv w_k \quad \forall j, k \in M$. Let $\mathcal{I}_{n \times m} \subseteq \mathcal{S}_{n \times m}$ be defined by

$$\mathcal{I}_{n \times m} \triangleq \{F \in \mathcal{S}_{n \times m} | \text{ F is facility symmetric }\}.$$

Define $\mathcal{IN}_{n \times m} \subseteq \mathcal{I}_{n \times m}$ as follows:

$\mathcal{IN}_{n \times m} \triangleq$
$\{F \in \mathcal{I}_{n \times m}|$ all facilities of F are non-increasing$\}$.

### 4.1 The Mediation Value

Although congestion games are especially interesting when the number of players is large, we first start with some results for the case where we have only two players, extending upon the results in the previous section. Following that, we will consider the more general $n$-player case.

#### 4.1.1 The two-player case ($n = 2$)

In theorem 3.1 we showed that $\frac{4}{3}$ is a tight upper bound for the mediation value of games that belong to $\mathcal{G}_{2 \times 2}$. Hence, obviously, $\frac{4}{3}$ is an upper bound for the mediation value of simple congestion games with two players and two facilities, i.e. games generated by congestion forms in $\mathcal{S}_{2 \times 2}$. We first show that this is also a tight upper bound.

**Theorem 4.1** $MV(\{\Gamma_F | F \in \mathcal{S}_{2 \times 2}\}) = \frac{4}{3}$.

Consider now the more general case where, the two agents can choose among $m$ facilities. We show:

**Theorem 4.2** $MV(\{\Gamma_F | F \in \mathcal{S}_{2 \times m}\}) \leq 2$.

Notice that our results imply that correlation helps already when we have congestion games with only two players. However, correlation does not increase social welfare when all facility payoff functions are non-increasing:

**Theorem 4.3** $MV(\Gamma_F) = 1$ for every form $F \in \mathcal{SN}_{2 \times m}$.

#### 4.1.2 Simple congestion games with $n$ players

In Section 3 we have shown that correlation has an unbounded value when considering arbitrary games. We next consider the effects of correlation in the context of simple congestion games. We can show:

**Theorem 4.4** $MV(\{\Gamma_F | F \in \mathcal{SN}_{n \times m}\}) = \infty$ for every $n \geq 3$ and for every $m \geq 2$.

The above result illustrates the power of correlation when we consider the context of simple congestion games. Indeed, the result shows that even if there are three players and two facilities with non-increasing payoff functions the mediation value is unbounded. However, if we require that the facility payoff functions are linear, then the following upper bound can be obtained:

**Theorem 4.5** $sup\{MV(\Gamma_F) | F \in \mathcal{SN}_{n \times 2},$
all facilities of F are linear$\} \leq \phi$, where $\phi = (\sqrt{5} + 1)/2$.

Proving that $\phi$ is an upper bound is highly non-trivial. Unfortunately, we do not know what is the least upper bound. However, the example below shows that the mediation value can be at least $\frac{9}{8}$.

**Example 1:** Let $n = 3$, $M = \{f, g\}$, $w_f = (24, 12, 0)$ and $w_g = (8, 8, 8)$. It can be shown that $v_N(\Gamma) = 32$, and it can be obtained, both in a pure-strategy equilibrium (two players choose $f$ and the other player chooses $g$) and in a mixed-strategy equilibrium. Consider the following correlated strategy $\mu$. Assign the probability $\frac{1}{6}$ to each strategy profile in which, not all players choose the same facility. $\mu$ is in equilibrium and the surplus at $\mu$ is 36. □

The above study shows that correlation is extremely helpful in the context of (even non-increasing) congestion games. We next show that correlation is helpful even in the narrow class of facility symmetric forms with non-increasing facilities.

**Theorem 4.6** $MV(\{\Gamma_F | F \in \mathcal{IN}_{n \times 2}\}) > 1$ for every $n \geq 4$.

The case of symmetric forms with non-increasing facilities is quite restricting one. As a result, the fact the mediation value may be greater than 1 in this case is quite encouraging. However, if we further restrict the setting to obey some concavity requirements, the above does not hold any more. Formally, we say that a function $v : \{1, 2, ..., N\} \to \mathbb{R}^+$ is *concave* if for every integer $k \geq 2$  $v(k+1) - v(k) \leq v(k) - v(k-1)$. We can now show:

**Theorem 4.7** Let $F \in \mathcal{IN}_{n \times m}$ and assume $n \geq m$. Define $v$ by $v(k) = kg(k)$. If $v$ is concave then there exists an equilibrium in $\Gamma_F$ which obtains the maximal surplus.

### 4.2 The Enforcement Value

We already know that the enforcement value is unbounded on the class of Prisoner's Dilemma games (notice that a Prisoner's Dilemma game is a simple congestion game). In addition, we show:

**Theorem 4.8** $sup\{EV(\Gamma) | F \in \mathcal{SN}_{3 \times 2},$
there are no strictly dominant strategies$\} = \infty$.

The above result shows that the enforcement value may be unbounded already on the class of simple congestion games with non-increasing facility payoff functions. It turns out that it is unbounded even when restricting the games to those who are generated by

symmetric congestion forms with non-increasing facility payoff functions:

**Theorem 4.9** $\lim_{n \to \infty} EV(\{\Gamma_F | F \in \mathcal{IN}_{n \times 2}\}) = \infty$.

Although the enforcement value may be unbounded when we have facility symmetric congestion forms, it is of great interest to characterize general cases of this natural setup where the correlation enables to get close to the maximal value. More specifically, we now characterize the cases where correlation allows to actually obtain the related maximal value, i.e. the enforcement value is 1.

The following characterization makes use of the following definition and notations. Let $F$ be a simple congestion form with $n$ players and $m$ facilities. A congestion vector $\pi = \pi(n,m)$ is an $m$-tuple $\pi = (\pi_j)_{j \in M}$, where $\pi_1, \pi_2, ..., \pi_m \in Z^*$ (nonnegative integers) and $\sum_{j=1}^{m} \pi_j = n$. $\pi$ represents the situation where $\pi_j$ players choose facility $j$. Every strategy profile $A \in X$ uniquely determines a congestion vector $\pi^A$. Note that there are $\binom{n}{\pi_1}\binom{n-\pi_1}{\pi_2} \cdots \binom{n-\sum_{j=1}^{m-2} \pi_j}{\pi_{m-1}}$ strategy profiles in the game $\Gamma_F$ that correspond to a congestion vector $\pi$, and denote by $B_\pi$ the set of all such strategy profiles. Thus $B_\pi = \{A \in X | \pi^A = \pi\}$. Given a congestion vector $\pi$, all strategy profiles in $B_\pi$ have the same surplus which we denote by $u(\pi)$. Therefore $u(\pi) = \sum_{j \in M} \pi_j w_j(\pi_j)$ where $w_j(0)$ is defined to be zero for every $j \in M$. We will say that a congestion vector $\pi$ is in equilibrium if every strategy profile in $B_\pi$ is in equilibrium. Let $\tau : M \to M$ be a one to one function and let $\tau\pi = (\tau\pi)_{j \in M}$ be the congestion vector defined by $(\tau\pi)_j = \pi_{\tau(j)}$. Let $F \in \mathcal{I}_{n \times m}$. In this case $u(\pi) = u(\tau\pi)$. Let $A_\pi = \bigcup_\tau B_{\tau\pi}$. Observe that $B_\pi \subseteq A_\pi$. Both sets are finite and therefore their elements can be ordered.

**Theorem 4.10** Let $F \in \mathcal{I}_{n \times m}$. Then $v_C(\Gamma_F) = opt(\Gamma_F)$ if and only if there exist a congestion vector $\pi = (\pi_1, ..., \pi_m)$ and a correlated equilibrium $\mu \in C(\Gamma_F)$ such that:

1. $u(\pi) = opt(\Gamma_F)$.

2. $\mu$ is distributed uniformly over all elements (strategy profiles) in $A_\pi$.

The proof of the above result show implicitly the existence of many situations where correlation allows to obtain the maximal surplus, while (without correlation) the best mixed-strategy equilibrium behaves poorly. We demonstrate this by the following example.

**Example 2:** Let $F \in \mathcal{I}_{6 \times 2}$. Let $w_j = (1.5, 1, 4, 4.5, 4.5, 3)$ for every $j \in M$. The maximal surplus is obtained in any strategy profile that belongs to $A_{\pi_1}$ and $A_{\pi_2}$ where $\pi_1 = (3,3)$ and $\pi_2 = (1,5)$. It is easy to see that the both $\pi_1$ and $\pi_2$ are not in equilibrium. Let $\xi_1$ and $\xi_2$ be correlated strategies that are uniformly distributed over $A_{\pi_1}$ and $A_{\pi_2}$ respectively. In order to check if there exists a correlated equilibrium that obtains the maximal surplus it is enough to check whether $\xi_1$ or $\xi_2$ are correlated equilibria. Indeed, one can easily check that $\xi_2$ is a correlated equilibrium whose surplus equals the maximal surplus. □

## Appendix

In this section we give a sketch of proof for Theorem 4.5, for illustrative purposes. One of the tools we use is linear programming. It is well-known that for every game in strategic form, $\Gamma$, $C(\Gamma)$ is exactly the set of feasible solutions for the following linear program $(\widehat{P})$. Moreover, $\mu \in C(\Gamma)$ is an optimal solution for $(\widehat{P})$ if and only if $u(\mu) = v_C(\Gamma)$.

$\max \sum_{s \in S} \mu(s) u(s)$

$\widehat{P}$ s.t.

$\mu(s) \geq 0 \quad \forall s \in S,$

$\sum_{s \in S} \mu(s) = 1,$

$\sum_{s^{-i} \in S^{-i}} \mu(s)[u^i(t^i, s^{-i}) - u^i(s)] \leq 0 \, \forall i \in N, \forall s^i \in S^i, \forall t^i \in S^i, t^i \neq s^i$

The dual problem has one decision variable for each constraint in the primal. We let $\alpha^i(t^i|s^i)$ denote the dual variable associated with the primal constraint:

$$\sum_{s^{-i} \in S^{-i}} \mu(s)[u^i(t^i, s^{-i}) - u^i(s)] \leq 0.$$

Let $\beta$ denote the dual variable associated with the primal constraint $\sum_{s \in S} \mu(s) = 1$. Let $\alpha = (\alpha^i)_{i \in N}$ where $\alpha^i = (\alpha^i(t^i|s^i))_{t^i, s^i \in S^i}$, $t^i \neq s^i$. The dual problem may be written:

$\min \beta$

$\widehat{D}$ s.t.

$\alpha^i(t^i|s^i) \geq 0 \quad \forall i \in N, \forall s^i \in S^i, \forall t^i \in S^i, t^i \neq s^i,$

$\sum_{i \in N} \sum_{s^i \neq t^i \in S^i} \alpha^i(t^i|s^i)[u^i(t^i, s^{-i}) - u^i(s)] + \beta \geq u(s) \forall s \in S$

It is well known that problems $\widehat{P}$ and $\widehat{D}$ are feasible and bounded, and their objective values equal $v_C(\Gamma)$.

**Proof of theorem 4.5(sketch):**
Let $M = \{f, g\}$ and let $w_f$ and $w_g$ be the facility payoff functions of $f$ and $g$ respectively. W.l.o.g $w_f(1) \geq w_g(1)$. Let $d_f = w_f(k) - w_f(k+1)$ and $d_g = w_g(k) - w_g(k+1)$ for every $1 \leq k \leq n-1$.

Let $\pi_k = (n - k, k)$ be the congestion vector where $k$ players choose $g$ and $n - k$ players choose $f$. Let $s$ be the largest integer such that the congestion vector $\pi_s = (n - s, s)$ is in equilibrium (a pure strategy equilibrium exists (Rosenthal, 1973)). The surplus in the above equilibrium is $u(\pi_s)$. If $s \in \{n, 0\}$ then the mediation value is one. We prove the following two claims:

**Claim 1**: $u(\pi_j) \leq u(\pi_s)$ for every $j \leq s$.

**Claim 2**: For every $k > s$ every strategy profile in the following form is in equilibrium: $n - k$ players choose $f$ with probability one, and the other $k$ players choose $g$ with probability $p_k = \frac{w_g(1) - w_f(n)}{(k-1)(d_f + d_g)}$. The surplus of such a strategy profile is $nw_f(n) + p_k d_f((n - k)k + k(k - 1))$.

We continue: let $q_{s+1}$ be a strategy profile such as in claim 2. In order to prove the theorem we use the dual program $\hat{D}$. Let $(\alpha, \beta)$ be a feasible solution for the dual problem, then by the duality theorem $\beta \geq v_C(\Gamma)$. Let $Z = \phi \max\{u(\pi_s), u(q_{s+1})\}$. We show that there exist a feasible solution for the dual problem where $\beta \leq Z$.

Let $x = \alpha^i(f|g)$ and $\alpha^i(g|f) = 0$ for every $i \in N$. The constraints of the dual program reduce to (call this system $\widehat{D_1}$):

$\widehat{D_1}$  $k(w_f(n - k + 1) - w_g(k))x \geq u(\pi_k) - \beta$,  $k = 1, ..., n$

$\quad x \geq 0$

Consider the case where there is at least one $k$ such that $u(\pi_k) > Z$. Let $\hat{k}$ be such that $u(\pi_{\hat{k}}) > Z$. Thus $\hat{k} > s$ (by claim 1). If there exists a feasible solution to $\widehat{D_1}$ where $\beta \leq Z$, it must be that $x \geq 0$. Therefore we can remove the constraint $x \geq 0$ from $\widehat{D_1}$. Call the new set of constraints (without $x \geq 0$) $\widehat{D_2}$.

By Farkas lemma, $\widehat{D_2}$ has a solution if and only if the following program doesn't have a solution:

$\widehat{P_1}$  $\sum_{k=1}^{n} y_k k(w_f(n - k + 1) - w_g(k)) = 0$

$\quad \sum_{k=1}^{n} y_k(u(\pi_k) - \beta) > 0$

$\quad y_k \geq 0 \quad k = 1, ..., n$

W.l.o.g let $y = (y_1, ..., y_n)$ be a probability distribution. Let $Y$ be the random variable where $y_k = P(Y = k) \quad k = 1, ..., n$. Suppose that there exist a vector $y$ that satisfies the first constraint. From the first constraint we obtain:

$$EY \leq \frac{w_g(1) - w_f(n) + d_f + d_g}{(d_f + d_g)} \leq s + 1 \quad (4)$$

We proceed with the second constraint. It remains to show that $\sum_{k=1}^{n} y_k u(\pi_k) \leq \beta$ for any $\beta \leq Z$. We obtain that:

$$\sum_{k=1}^{n} y_k u(\pi_k) = EY(w_f(1) - w_f(n)) + nw_f(n). \quad (5)$$

We distinguish between two cases: $(i)$ $p_{s+1} < 1/\phi$. $(ii)$ $p_{s+1} \geq 1/\phi$.
$(i)$ $p_{s+1} < 1/\phi$ implies that $(w_g(1) - w_f(n))/(d_f + d_g) \leq s/\phi$. From (4) $EY \leq \frac{s}{\phi} + 1$. Let $\beta = Mu(\pi_s)$ where $M \geq 1$. We show that if the second constraint is satisfied then every $M \geq \phi$ will contradict that $EY \leq \frac{s}{\phi} + 1$. $(ii)$ Let $p_{s+1} \geq 1/\phi$. From (4) we have $EY \leq s + 1$. Let $\beta = Mu(q_{s+1})$ where $M \geq 1$. We show that if the second constraint is satisfied then every $M \geq \phi$ contradicts that $EY \leq s + 1$. □